%% file: main.tex
\documentclass[11pt,a4paper]{article}
\usepackage[margin=2.45cm]{geometry}
\usepackage{amsmath,amssymb,mathtools,bm,graphicx}
\usepackage[T1]{fontenc}
\usepackage{flafter}
\usepackage{placeins}
\usepackage{float}
\usepackage[colorlinks=true,linkcolor=blue,citecolor=blue,urlcolor=blue]{hyperref}
\newcommand{\LN}{\mathcal{L}_{\mathcal N}}
\newcommand{\Tr}{\operatorname{Tr}}
\newcommand{\Spec}{\operatorname{spec}}
\newcommand{\HH}{\mathsf H}

\title{Pairwise fermionic mode correlations in electric and magnetic backgrounds:\\ Landau-level structure and Schwinger production}

\author{%
Sahil Chetan Patnekar\textsuperscript{1},
Sristi Sanjivanee Goswami\textsuperscript{1},
Sanjay Pant\textsuperscript{2}, and
Shagun Kaushal\textsuperscript{1,*}\\[1ex]
\small \textsuperscript{1}\textit{Department of Physics,
School of Advanced Sciences,}\\
\small \textit{Vellore Institute of Technology,
Vellore 632014, India}\\[0.4ex]
\small \textsuperscript{2}\textit{Department of Physics,
School of Physical Sciences,}\\
\small \textit{DIT University, Mussoorie-Diversion Road,
Makkawala, Dehradun, Uttarakhand 248009, India}
}

\date{}

\begin{document}

\maketitle

\begingroup
\renewcommand{\thefootnote}{\fnsymbol{footnote}}
\setcounter{footnote}{1}
\footnotetext{Corresponding author:
\href{mailto:shagun.kaushal@vit.ac.in}
{shagun.kaushal@vit.ac.in}}
\endgroup

\setcounter{footnote}{0}

\begin{abstract}

We investigate mutual information and logarithmic
negativity in fermionic Schwinger pair production
in parallel electric and magnetic fields. Using
spin-dependent Bogoliubov transformations, we
construct the in-vacuum and derive the
particle--antiparticle correlations for selected
Landau modes. We also consider an initially
entangled two-particle state and evaluate the
correlations for six bipartitions. The physical
Landau-level structure is treated explicitly by
distinguishing modes with the same oscillator
index from the spin representatives of a common
physical Landau level. For matched higher levels,
the pairwise redistribution relations retain the
electric-only form, with the magnetic-field
dependence entering through the Landau-level
pair-production probabilities. When the two
particle groups occupy different physical levels,
the cross particle--antiparticle correlations
become unequal, and we derive analytic relations
characterizing this asymmetry. We further obtain
the local logarithmic negativity from the ordinary
partial transpose in a fixed occupation-space
basis. Finally, we extend the analysis to a
Sauter-type electric field in a constant magnetic
background and examine the dependence of the
pairwise correlations on the field parameters
and Landau-level structure.
\end{abstract}
\input{paper_body.tex}

\end{document}

%% file: paper_body.tex
\section{Introduction}
The creation of charged particle--antiparticle pairs by an external electric field is a non-perturbative effect in quantum field theory, known as the Schwinger effect \cite{Sauter1931,HeisenbergEuler1936,Schwinger1951,Dunne2005,DiPiazza2012}. The in- and out-vacua are related through Bogoliubov transformations, and the produced particle and antiparticle modes are consequently correlated. For fermions, the occupation number of an elementary particle--antiparticle mode pair is restricted by the Pauli exclusion principle, allowing its reduced density matrix to be evaluated explicitly \cite{Ebadi2014,PlenioVirmani2007}.

The Schwinger effect has been investigated in time-dependent and spatially inhomogeneous backgrounds, using both field-theoretic and quantum-kinetic approaches \cite{Cooper1993,GiesKling2005,Schutzhold2008,DunneGies2009,DumluDunne2010,Hebenstreit2011,Fillion2013,Gonoskov2013,Blinne2014,Jansen2016,Wollert2016,KimLeeYoon2008,Gavrilov2015}. Quantum correlations of produced modes have received considerable attention \cite{Fuentes2010,MartinMartinezFuentes2011,WuZeng2020}. Ebadi and Mirza considered the generation of entanglement in electric backgrounds \cite{Ebadi2014}; subsequently, Li, Dai and Shi studied the redistribution of mutual information and logarithmic negativity among different particle--antiparticle pairs for vacuum and initially entangled states \cite{Li2017}. Related questions have been explored for quantum fields in noninertial and curved backgrounds \cite{Shi2004,FuentesSchuller2005,Leon2009,MartinLeon2010,Dai2015,Dai2016,Ball2006,Moradi2014,Parker1968,BirrellDavies1982,Hawking1975}.

In the presence of a magnetic field, the transverse
motion of charged fermions is quantized into Landau
levels, modifying the effective transverse energy
that enters the Schwinger pair-production probability.
For the flat-spacetime Dirac field, the spin-resolved modes
and Bogoliubov transformations in parallel electric and
magnetic backgrounds are developed in
Ref.~\cite{Kaushal2025}. Related studies of
Schwinger correlations in scalar, curved-spacetime,
and accelerated settings can be found in
Refs.~\cite{Bhattacharya2020,Ali2021,Kaushal2022}.
In Ref.~\cite{Kaushal2025}, fermionic entanglement
was studied in parallel electric and magnetic
backgrounds, including a Sauter-type time-dependent electric field.
The correlations were characterized using entanglement
entropy, mutual information, and Bell's inequality
violation for selected vacuum and initially entangled
states. Related questions have also been considered
in curved-space and holographic Schwinger settings
\cite{Ebadi2015,Grieninger2023a,Grieninger2023b}.
More recently, logarithmic negativity associated
with Schwinger particle production has been
investigated for scalar and spinor fields, including
thermal initial states \cite{Kranas2026}.
The charged-scalar Schwinger--Unruh problem in Rindler spacetime has also been examined in Ref.~\cite{Kaushal:2026ctx}. These studies provide the basis for a more detailed
examination of pairwise correlations in
Landau-quantized fermionic modes.

In the present work, we extend the pairwise-mode
analysis of Ref.~\cite{Li2017} to fermionic
Schwinger production in parallel electric and
magnetic fields, using the spin-resolved
Bogoliubov transformations of Ref.~\cite{Kaushal2025}.
We first construct the spin-dependent in-vacuum
and calculate the mutual information and
logarithmic negativity between particle and
antiparticle modes. We then consider an initially
entangled two-particle state and obtain the
correlations for six particle--antiparticle
bipartitions. Particular attention is given to
the distinction between states having the same
oscillator index and the degenerate spin
representatives of a common physical Landau level.
For matched higher Landau levels, we examine the
pairwise correlation-redistribution relations
and their dependence on the effective transverse
gap. For particle groups occupying different physical
Landau levels, we derive analytic relations
characterizing the asymmetry between the two
cross particle--antiparticle correlations.
Unlike the matched-level case, these correlations
are generally unequal because of the different
transverse gaps. We also obtain the local
particle--antiparticle logarithmic negativity
from the reduced density-matrix spectrum
using the ordinary partial transpose.
Finally, we extend the vacuum and initially
entangled state calculations to a Sauter-type
electric field in the presence of a constant
magnetic background.

The remainder of this paper is organized as follows.
In Sec.~\ref{sec:setup}, we introduce the theoretical setup
and the correlation measures used in our analysis.
In Sec.~\ref{sec:vacuum}, we investigate the mutual information
and logarithmic negativity for the spin-dependent vacuum state.
In Sec.~\ref{sec:entangled}, we consider an initially entangled
particle state and examine the redistribution of correlations
among different particle--antiparticle bipartitions.
In Sec.~\ref{sec:sauter}, we extend our analysis to a
Sauter-type electric field in the presence of a background
magnetic field. Finally, we summarize our main results
in Sec.~\ref{sec:conclusion}.
Throughout this work, we use natural units
$\hbar=c=k_B=1$ and adopt the mostly positive
metric signature.

\section{The setup}\label{sec:setup}
\subsection{Dirac fermions in constant parallel fields}
We consider a Dirac field of mass $m$ and charge $e$ in the presence of parallel electric and magnetic fields, $\mathbf E=E\hat{\mathbf z}$ and $\mathbf B=B\hat{\mathbf z}$. Its equation of motion is
\begin{equation}
\left(i\gamma^\mu\partial_\mu-e\gamma^\mu A_\mu-m\right)\Psi(x)=0.
\label{eq:dirac}
\end{equation}
In the static gauge used in Ref.~\cite{Kaushal2025}, $A_\mu=(-Ez,-By,0,0)$. Introducing a second-order spinor variable and separating the transverse motion gives harmonic-oscillator wave functions. The separation constants in the spin convention of that reference are
\begin{equation}
S_{n,\uparrow}=2(n+1)|eB|,\qquad S_{n,\downarrow}=2n|eB|,\qquad n=0,1,2,\ldots.
\label{eq:S}
\end{equation}
The longitudinal equations reduce to parabolic-cylinder equations. The spinor mode solutions, normalization and operator relations are discussed in detail in Ref.~\cite{Kaushal2025}; we summarize the properties needed for the correlation calculation.

Let $a_{r,s}$ denote the particle annihilator in an orthonormal Dirac mode and $b_{\bar r,\bar s}$ the annihilator of its paired antiparticle mode. Here $\bar s$ labels the \emph{Bogoliubov partner}; it need not represent the same physical spin projection as $s$. In particular, the mode index $s$ used to label the spinor basis in Ref.~\cite{Kaushal2025} must not be identified with a laboratory antiparticle spin projection without specifying the charge-conjugation convention. In the real-amplitude phase convention used below, the canonical relations for each independent particle--antiparticle mode pair are
\begin{align}
a^{\mathrm{in}}_{r,s}&=\sqrt{A_{r,s}}\,a^{\mathrm{out}}_{r,s}-\sqrt{P_{r,s}}\,b^{\mathrm{out}\dagger}_{\bar r,\bar s},\\
b^{\mathrm{in}}_{\bar r,\bar s}&=\sqrt{A_{r,s}}\,b^{\mathrm{out}}_{\bar r,\bar s}+\sqrt{P_{r,s}}\,a^{\mathrm{out}\dagger}_{r,s}.
\label{eq:operatorpair}
\end{align}
A rephasing of the out operators changes the displayed Bogoliubov phases without changing $P_{r,s}=|\beta_{r,s}|^2$ or the correlation measures considered here. The fermionic anticommutation relations imply $A_{r,s}+P_{r,s}=1$. The source-mode overlap is diagonal in the mode index for the normalized basis of Ref.~\cite{Kaushal2025}; the derivation of the associated occupation-space map and its signs is given in Appendix~\ref{app:operatorpair}. For constant fields, the probabilities read
\begin{align}
P_{n,\uparrow}\equiv|\beta_{n,\uparrow}|^2&=\exp\left[-\frac{\pi(m^2+2(n+1)|eB|)}{|eE|}\right],\\
P_{n,\downarrow}\equiv|\beta_{n,\downarrow}|^2&=\exp\left[-\frac{\pi(m^2+2n|eB|)}{|eE|}\right],\qquad A_{n,s}\equiv 1-P_{n,s}=|\alpha_{n,s}|^2.
\label{eq:const_probs}
\end{align}
We retain the full spin labels in this oscillator-index parameterization. At the same oscillator index, $P_{n,\uparrow}<P_{n,\downarrow}$ when $B>0$; these states, however, belong to \emph{different physical Landau-energy levels}. In particular, with $N$ denoting the physical Landau level,
\begin{equation}
(n=N-1,\uparrow)\ \longleftrightarrow\ (n=N,\downarrow),\qquad N\geq1,
\label{eq:physicalmatching}
\end{equation}
and both modes have $S=2N|eB|$. Therefore, at fixed longitudinal momentum and other quantum numbers,
\begin{equation}
P_{N-1,\uparrow}=P_{N,\downarrow}\equiv P_N=\exp\left[-\frac{\pi(m^2+2N|eB|)}{|eE|}\right],\quad A_N=1-P_N.
\label{eq:physicalP}
\end{equation}
Thus the same-$n$ comparison in our figures compares two different Landau energies, not a lifted spin degeneracy within one physical level. The only independent mode pair at $N=0$ is $(n=0,\downarrow)$ in the chosen convention; the higher levels have two spin representatives. These are properties of the ideal minimally coupled Dirac spectrum, and are not modified by the external electric field in the present setup \cite{Bhattacharya2020, Kaushal2025}. Accordingly, a two-mode-pair state at a fixed oscillator index is an algebraic sector model, and the lowest-level physical contribution is evaluated separately. Results for individual fixed Landau modes should also not be identified with total correlations summed over the transverse degeneracy. In particular, setting $B=0$ at fixed $n$ tests equal-spin algebra but does not implement the full continuum limit of the Landau sum.

\subsection{Correlation measures}
For a bipartite density matrix $\rho_{AB}$ we define \cite{PlenioVirmani2007, NielsenChuang2000}
\begin{equation}
I(A:B)=S(\rho_A)+S(\rho_B)-S(\rho_{AB}),\qquad S(\rho)=-\Tr(\rho\log_2\rho),
\label{eq:MI}
\end{equation}
where $\rho_A=\Tr_B\rho_{AB}$ and $\rho_B=\Tr_A\rho_{AB}$. The mutual information is a measure of total correlations, including classical and quantum contributions. By contrast, logarithmic negativity detects negative eigenvalues of the partial transpose \cite{VidalWerner2002,Plenio2005},
\begin{equation}
\LN(A:B)=\log_2\|\rho_{AB}^{T_A}\|_1=\log_2\sum_j|\lambda_j(\rho_{AB}^{T_A})|.
\label{eq:LNdef}
\end{equation}
Throughout this work, logarithmic negativity is evaluated using the ordinary partial transpose in a fixed occupation-space basis, following Ref.~\cite{Li2017}. The fermionic anticommutation signs are retained when constructing the occupation states, and the same operator ordering and subsystem partitions are used throughout the calculations.

For mixed states, separability and entanglement criteria must be distinguished from entropy-based measures \cite{Horodecki1996,Zyczkowski1998,Bruss2002}. Related developments include entanglement monotones and their monogamy constraints \cite{Coffman2000,Osborne2006}, extensions of logarithmic negativity \cite{WangWilde2020}, and field-theoretic treatments of entropy and negativity \cite{Calabrese2013,Nishioka2018}.

For later use, the binary entropy is
\begin{equation}
h_2(x)=-x\log_2x-(1-x)\log_2(1-x),\qquad 0\log_2 0\equiv0.
\label{eq:h}
\end{equation}

\section{Vacuum state and results}\label{sec:vacuum}
\subsection{Spin-dependent vacuum and quantum correlations}
With the antiparticle arrows understood as labels of the paired antiparticle slots in the convention of Ref.~\cite{Li2017} (not as an independent measurement of the antiparticle spin along $B$), the spin-dependent in-vacuum can be expressed in the out basis as
\begin{align}
|0_k,0_{-k}\rangle_{\rm in}={}&x_0|0_k,0_{-k}\rangle_{\rm out}+x_1|\uparrow_k,\downarrow_{-k}\rangle_{\rm out}
+x_2|\downarrow_k,\uparrow_{-k}\rangle_{\rm out}
+x_3|\uparrow\downarrow_k,\uparrow\downarrow_{-k}\rangle_{\rm out}.
\label{eq:vacstate}
\end{align}
The coefficients, including their relative phases, can be obtained by imposing annihilation of the in-vacuum by the in-particle and in-antiparticle operators. For the two independent particle--antiparticle mode pairs, their squared magnitudes are
\begin{equation}
|x_0|^2=A_\uparrow A_\downarrow,\quad |x_1|^2=P_\uparrow A_\downarrow,\quad |x_2|^2=A_\uparrow P_\downarrow,\quad |x_3|^2=P_\uparrow P_\downarrow.
\label{eq:x}
\end{equation}
Here, all $A_s$ and $P_s$ refer to the selected Landau modes. Directly, $\sum_j|x_j|^2=(A_\uparrow+P_\uparrow)(A_\downarrow+P_\downarrow)=1$. The phases do not enter the two vacuum correlation measures considered below.

In the four-dimensional correlated support of Eq.~\eqref{eq:vacstate}, the complete density matrix is $\rho_{k,-k}=xx^\dagger$ with $x=(x_0,x_1,x_2,x_3)^T$, or explicitly
\begin{equation}
\rho_{k,-k}=\begin{pmatrix}
|x_0|^2&x_0x_1^*&x_0x_2^*&x_0x_3^*\\
x_1x_0^*&|x_1|^2&x_1x_2^*&x_1x_3^*\\
x_2x_0^*&x_2x_1^*&|x_2|^2&x_2x_3^*\\
x_3x_0^*&x_3x_1^*&x_3x_2^*&|x_3|^2
\end{pmatrix}.
\end{equation}
As it represents a normalized pure state, $\Tr\rho_{k,-k}^2=1$ and $S(\rho_{k,-k})=0$. Tracing out the antiparticle subsystem gives a diagonal, generally mixed reduced state,
\begin{equation}
\rho_k=\Tr_{-k}\rho_{k,-k}=\mathrm{diag}\left(A_\uparrow A_\downarrow,\ P_\uparrow A_\downarrow,\ A_\uparrow P_\downarrow,\ P_\uparrow P_\downarrow\right).
\label{eq:rhok}
\end{equation}
Its purity is $(A_\uparrow^2+P_\uparrow^2)(A_\downarrow^2+P_\downarrow^2)$ and is less than one if the relevant Schwinger probabilities lie strictly between zero and one.

The four eigenvalues of Eq.~\eqref{eq:rhok} are products of the two independent Bernoulli distributions. Therefore,
\begin{equation}
S(\rho_k)=h_2(P_\uparrow)+h_2(P_\downarrow).
\end{equation}
The full state being pure, $S(\rho_{-k})=S(\rho_k)$ and $S(\rho_{k,-k})=0$, which yields
\begin{equation}
{I_{\mathrm{vac}}(k:-k)=2h_2(P_\uparrow)+2h_2(P_\downarrow).}
\label{eq:Ivac}
\end{equation}
For a single physically independent spin-resolved mode, $I_s=2h_2(P_s)$. This is maximal at $P_s=1/2$, vanishes when $P_s\to0$ or $P_s\to1$, and remains nonzero for an intermediate occupation probability. Hence the mutual information need not grow monotonically with electric-field strength even though $P_s$ itself grows with $E$.

For a pure bipartite state with Schmidt amplitudes $|x_j|$, the ordinary partial transpose has eigenvalues $|x_j|^2$ and $\pm|x_ix_j|$ for $i<j$. Consequently,
\begin{equation}
\|\rho_{k,-k}^{T_k}\|_1=\left(\sum_{j=0}^3|x_j|\right)^2
=\left[(\sqrt{A_\uparrow}+\sqrt{P_\uparrow})(\sqrt{A_\downarrow}+\sqrt{P_\downarrow})\right]^2.
\end{equation}
Thus,
\begin{equation}
{\LN^{\mathrm{vac}}(k:-k)=2\sum_{s=\uparrow,\downarrow}\log_2\left(\sqrt{1-P_s}+\sqrt{P_s}\right).}
\label{eq:LNvac}
\end{equation}
Each independent spin-resolved mode contributes $\LN{}_{s}=2\log_2(\sqrt{1-P_s}+\sqrt{P_s})$. The maximum is one bit per mode, attained at $P_s=1/2$. For identical spin-resolved modes the expressions reduce to $I_{\rm vac}=4h_2(P)$ and $\LN^{\rm vac}=4\log_2(\sqrt{1-P}+\sqrt P)$, providing the equal-spin check against the electric-only calculation \cite{Li2017}.

It is useful to distinguish the entropy and negativity of a single spin-resolved Schwinger pair from the correlations of the complete momentum-mode groups. The same grouped-mode partition is adopted by Ref.~\cite{Li2017}. In their electric-only treatment, a momentum mode comprises the four occupation states $|0\rangle$, $|\uparrow\rangle$, $|\downarrow\rangle$ and $|\uparrow\downarrow\rangle$. Hence, the particle and antiparticle groups each have Hilbert-space dimension four. In particular, they found a maximum vacuum logarithmic negativity of $2$ at $|\alpha|^2=|\beta|^2=1/2$; see their discussion following the vacuum logarithmic-negativity expression. The corresponding maximum mutual information is $4$ because the complete vacuum is pure and its reduced particle density matrix is maximally mixed on a four-dimensional space.

For a single spin-resolved pair, $0\le I_s\le2$ and $0\le{\LN}_s\le1$. The complete two-mode-pair vacuum, formed by grouping both spins into one particle subsystem and both antiparticle partners into the other, instead satisfies
\begin{equation}
0\le I_{\mathrm{vac}}(k:-k)\le4,\qquad
0\le\LN^{\mathrm{vac}}(k:-k)\le2.
\label{eq:vacboundcomparison}
\end{equation}
The bounds are kinematic; simultaneous saturation requires $P_\uparrow=P_\downarrow=1/2$. When the two effective transverse gaps differ at fixed nonzero $B$, the maxima of the individual modes occur at distinct electric-field values and these upper bounds cannot generally be saturated simultaneously.

For $B=0$ at the level of equal spin-sector amplitudes, $P_\uparrow=P_\downarrow=P$, Eqs.~\eqref{eq:Ivac} and \eqref{eq:LNvac} reduce to
\begin{equation}
I_{\mathrm{vac}}=4h_2(P),\qquad
\LN^{\mathrm{vac}}=4\log_2(\sqrt{1-P}+\sqrt P),
\end{equation}
which agrees with the grouped-spin vacuum calculation of Ref.~\cite{Li2017}. In addition, the maxima $I=2$ and $\LN=1$ for the initially entangled particle--particle subsystem $p:q$ reported in Ref.~\cite{Li2017} follow from a different initial state and bipartition; they do not constrain the complete vacuum $(k:-k)$ correlations. Thus the different bounds follow from the dimensions of the selected subsystems.

\subsection{Spin dependence and physical Landau-level matching}\label{sec:matched}
The unequal transverse gaps in Eq.~\eqref{eq:S} imply $P_{n,\uparrow}<P_{n,\downarrow}$ when $B>0$ and the same oscillator index is used. This ordering does not imply an ordering of $I_s$ or $\LN{}_{s}$ for all fields, because both measures are symmetric under $P_s\leftrightarrow1-P_s$ and reach their maxima at $P_s=1/2$. The spin-down mode reaches $P_s=1/2$ when
\begin{equation}
\frac{|eE|}{m^2}=\frac{\pi}{\ln2}\left(1+2n\frac{|eB|}{m^2}\right),
\end{equation}
whereas the corresponding spin-up maximum occurs after replacing $n$ by $n+1$. Hence the two spin-resolved modes can have distinct correlation maxima as the electric and magnetic fields vary.

The comparison at fixed $n$ therefore concerns different transverse energies. We next distinguish it from the matching of spin representatives at a common physical Landau level.

The oscillator index $n$ and physical Landau-level index $N$ are not interchangeable. In the convention of Eq.~\eqref{eq:S}, the two spin representatives of a common higher physical level $N\geq1$ are $(n=N-1,\uparrow)$ and $(n=N,\downarrow)$. They have the same transverse gap $2N|eB|$ and hence the same production probability $P_N$, as in Eqs.~\eqref{eq:physicalmatching} and \eqref{eq:physicalP}. The unequal spin contributions at equal $n$ discussed above therefore represent different orbital energies, not spin splitting within a degenerate physical level. This gives the level-resolved grouped vacuum correlations
\begin{align}
I_N^{\mathrm{vac}}&=4h_2(P_N),\\
\mathcal{L}_{\mathcal N,N}^{\mathrm{vac}}&=4\log_2\left(\sqrt{1-P_N}+\sqrt{P_N}\right).
\label{eq:matchedvac}
\end{align}
The physical lowest Landau level $N=0$ contains only $(n=0,\downarrow)$, with $P_0=\exp(-\pi m^2/|eE|)$. It consequently has $I_0^{\mathrm{vac}}=2h_2(P_0)$ and $\mathcal{L}_{\mathcal N,0}^{\mathrm{vac}}=2\log_2(\sqrt{1-P_0}+\sqrt{P_0})$. These selected-mode values have no explicit $B$ dependence; the degeneracy of the physical Landau level does, but is not summed over here. The limit $B\to0$ at a fixed $N$ is likewise distinct from the transverse continuum limit.

\subsection{Constant-field numerical results}

\begin{figure}[H]
\setlength{\abovecaptionskip}{3pt}
\setlength{\belowcaptionskip}{0pt}
\centering
\includegraphics[width=.55\linewidth]{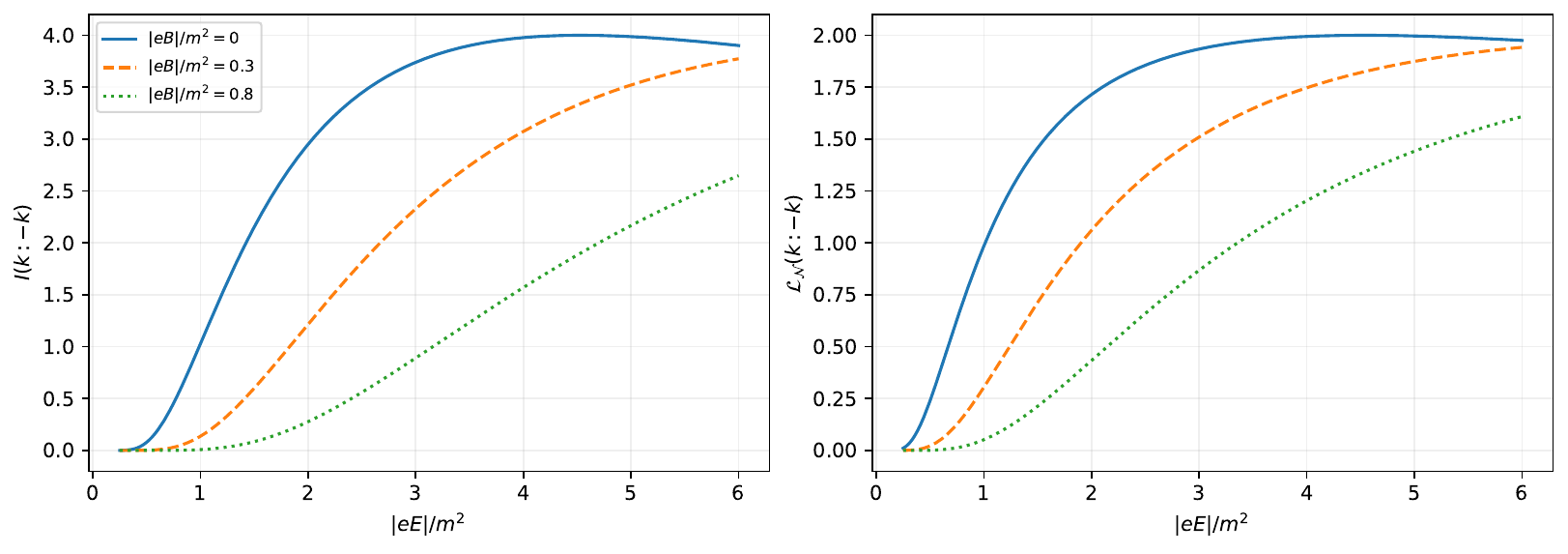}
\caption{\it \small Vacuum mutual information (left) and logarithmic negativity (right) as functions of $\mathcal E=|eE|/m^2$ for $\mathcal B=0$, $0.3$, and $0.8$ at fixed Landau oscillator index $n=1$. The in-vacuum is the four-term spin-dependent state of Eq.~\eqref{eq:vacstate}. The traces are evaluated for the grouped particle and antiparticle subsystems. Curves correspond to $I_{\rm vac}=2[h_2(P_\uparrow)+h_2(P_\downarrow)]$ and $\LN^{\rm vac}=\sum_s 2\log_2(\sqrt{1-P_s}+\sqrt{P_s})$.}
\label{fig:vacline}
\vspace{1mm}
\includegraphics[width=.55\linewidth]{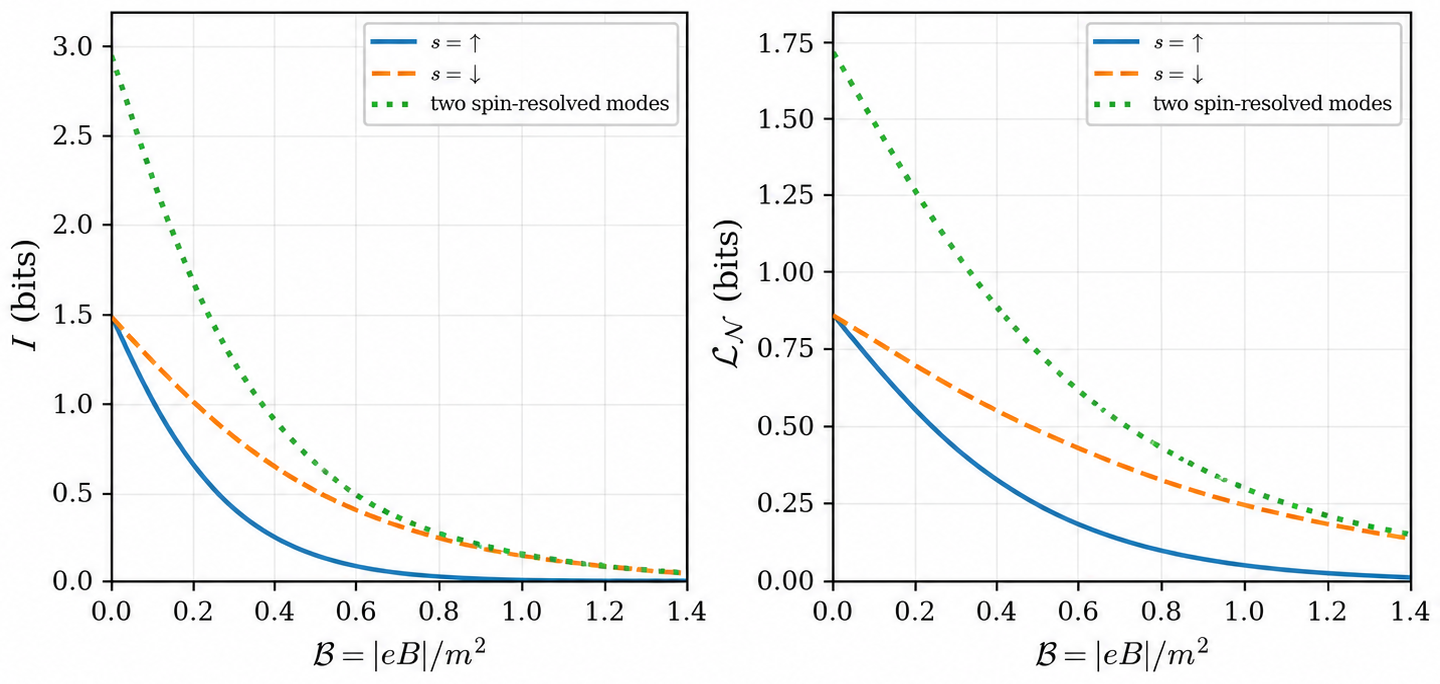}
\caption{\small \it Individual spin-up (solid) and spin-down (dashed) contributions to the vacuum mutual information (left) and logarithmic negativity (right) versus $\mathcal B=|eB|/m^2$, with $\mathcal E=2$ and fixed oscillator index $n=1$. The dotted curves give the sum of both independent spin contributions. The spin labels correspond to $S_{n,\uparrow}=2(n+1)|eB|$ and $S_{n,\downarrow}=2n|eB|$, and the two-mode-pair result is not a substitute for the single-mode physical lowest Landau level.}
\label{fig:vacspin}
\end{figure}

\begin{figure}[H]\centering\includegraphics[width=.82\linewidth]{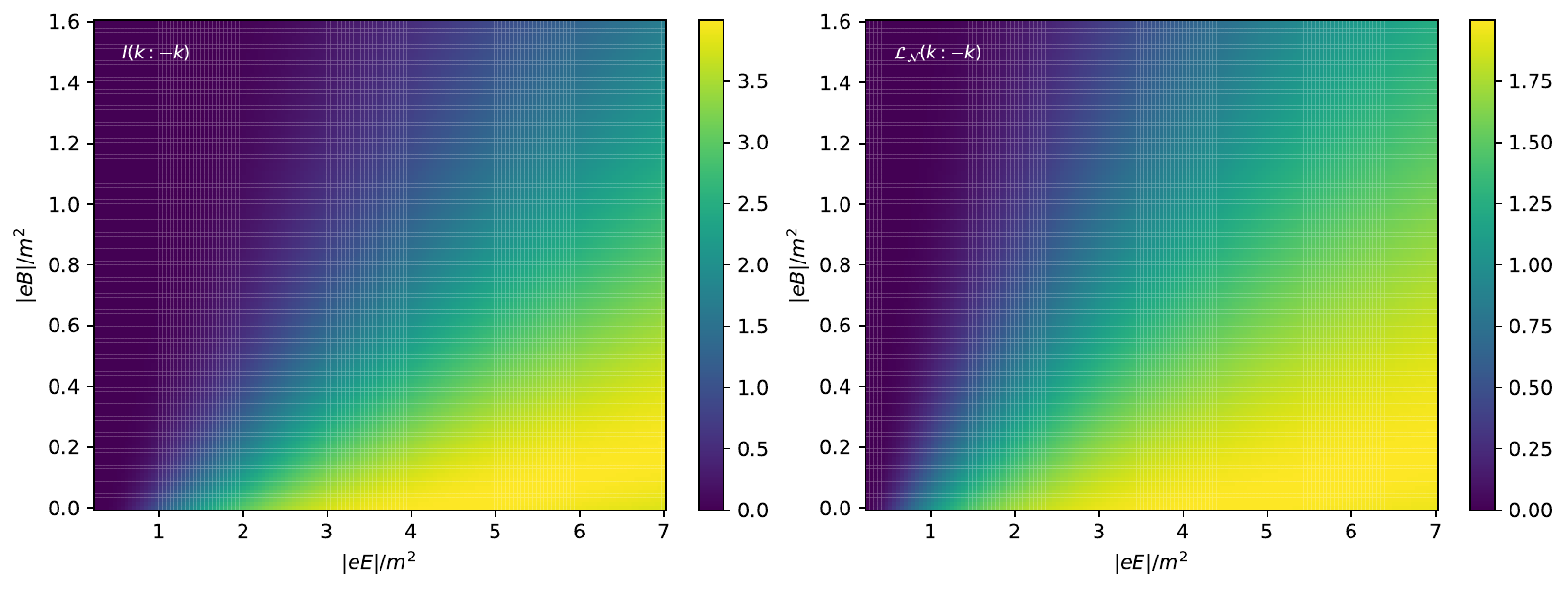}
\caption{\small \it Constant-field density maps of the vacuum mutual information (left) and logarithmic negativity (right) in the $(\mathcal E,\mathcal B)$ plane, with $\mathcal E=|eE|/m^2$, $\mathcal B=|eB|/m^2$, and $n=1$. Both maps use the same four-term spin-dependent in-vacuum; the color scales indicate the value of each measure in bits. No Landau-degeneracy sum is performed.}
\label{fig:vacdensity}\end{figure}

\begin{figure}[H]\centering\includegraphics[width=.80\linewidth]{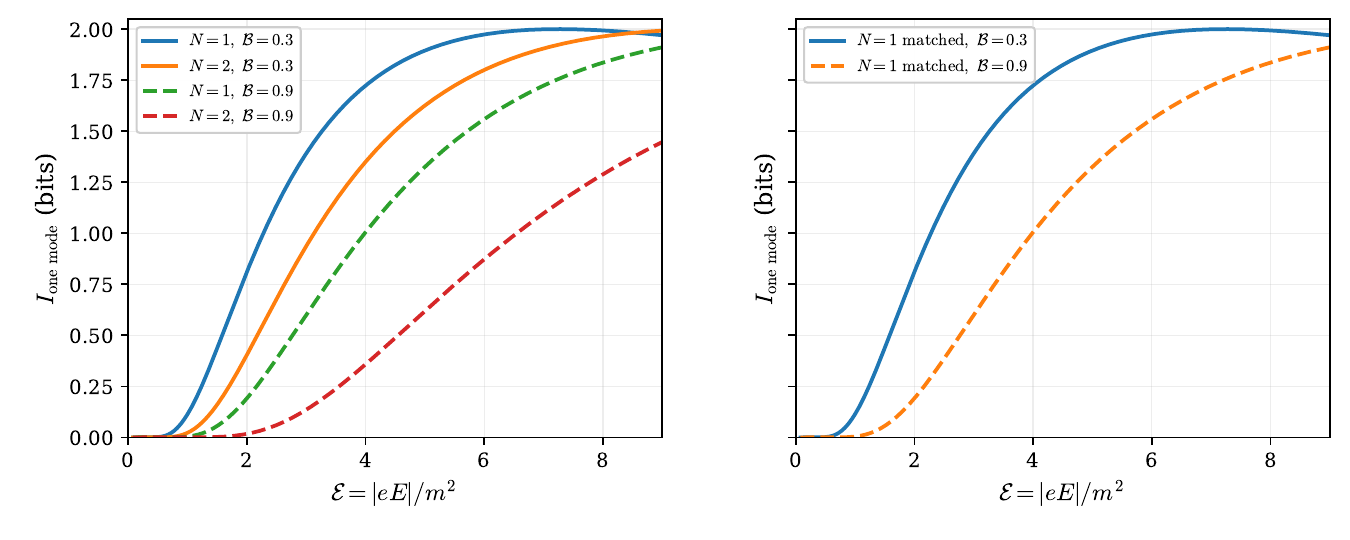}
\caption{\small \it Illustration of the Landau-index distinction in the single-mode vacuum mutual information, plotted versus $\mathcal E=|eE|/m^2$ at $\mathcal B=|eB|/m^2=0.3$ and $0.9$. The left panel shows single-mode mutual-information curves for $N=1$ and $N=2$: these correspond to the two branches when the orbital index is held fixed at $n=1$. The right panel shows the two degenerate spin representatives $(n=0,\uparrow)$ and $(n=1,\downarrow)$ of the same physical level $N=1$, which have coincident curves. Each curve is $I_{\mathrm{one\ mode}}=2h_2(P_N)$; no physical-level sum is performed.}\label{fig:levelcomparison}\end{figure}

We use $\mathcal E=|eE|/m^2$ and $\mathcal B=|eB|/m^2$. The fixed-$n$ illustrations take $n=1$, comparing physical levels $N=2$ and $N=1$. All results refer to selected modes. The field-dependent vacuum correlations are illustrated in Figs.~\ref{fig:vacline}--\ref{fig:vacdensity}. Figure~\ref{fig:levelcomparison} illustrates the distinction between equal oscillator indices and matched physical Landau levels.


\FloatBarrier
\section{Initially entangled state}
\label{sec:entangled}
\subsection{Initial state and reduced density matrices}
Following the two-particle construction of Ref.~\cite{Li2017}, we take
\begin{equation}
|\Phi\rangle_{\rm in}=\epsilon|\uparrow_p,\downarrow_q\rangle_{\rm in}+\sqrt{1-\epsilon^2}|\downarrow_p,\uparrow_q\rangle_{\rm in},\qquad0\le\epsilon\le1,
\label{eq:phi}
\end{equation}
with the corresponding antiparticle modes initially empty. The labels $p,q$ denote two selected particle-mode groups and $-p,-q$ their antiparticle partners. Each grouped mode contains two spin occupations. To obtain the final state from the spin-dependent Bogoliubov relations, it is convenient to choose a definite occupation ordering $(p\uparrow,p\downarrow,q\uparrow,q\downarrow)$ and the corresponding partner slots. With the same ordered two-mode convention as Eq.~\eqref{eq:operatorpair}, the elementary normalized map of a mode pair is
\begin{align}
|0\rangle_{\mathrm{in}}&\mapsto\sqrt{A_{r,s}}|00\rangle_{\mathrm{out}}+\sqrt{P_{r,s}}|11\rangle_{\mathrm{out}},\\
|1\rangle_{\mathrm{in}}&\mapsto|10\rangle_{\mathrm{out}}.
\label{eq:iso}
\end{align}
Here the two output occupations label the chosen particle mode and its paired antiparticle mode. This is an isometric occupation representation, $V_{r,s}^\dagger V_{r,s}=\mathbf1$, used consistently in the numerical implementation. Keeping all four mode pairs gives
\begin{equation}
|\Psi\rangle_{\mathrm{out}}=(V_{p,\uparrow}\otimes V_{p,\downarrow}\otimes V_{q,\uparrow}\otimes V_{q,\downarrow})\left[\epsilon|1,0,0,1\rangle+\sqrt{1-\epsilon^2}|0,1,1,0\rangle\right].
\label{eq:out}
\end{equation}
The probability amplitudes in the two brackets are not assumed equal unless $\epsilon=1/\sqrt2$. This construction includes Pauli blocking in modes that are initially occupied.

We study six grouped bipartitions,
\begin{equation}
(p:q),\ (p:-q),\ (-p:q),\ (-p:-q),\ (p:-p),\ (q:-q).
\label{eq:partitions}
\end{equation}
For any such pair $(X,Y)$, the reduced density matrix is obtained by tracing the other two groups: $\rho_{XY}=\Tr_{\overline{XY}}|\Psi\rangle\langle\Psi|$. In an ordered tensor-product basis, if $C_{abr}$ denotes the coefficient with kept indices $a,b$ and traced index $r$, then
\begin{equation}
(\rho_{XY})_{ab,a'b'}=\sum_r C_{abr}C_{a'b'r}^*,\quad
(\rho^{T_X}_{XY})_{ab,a'b'}=(\rho_{XY})_{a'b,ab'}.
\label{eq:traceformula}
\end{equation}
Equation~\eqref{eq:traceformula} is sufficient to reproduce every finite-dimensional density matrix and ordinary partial transpose from Eq.~\eqref{eq:out}. In the numerical calculation all grouped subsystems have dimension four, so that each pair density matrix is at most $16\times16$.

\subsection{Pairwise mutual information and logarithmic negativity}
For the $p:q$ sector, define $u=\epsilon^2$, $v=1-\epsilon^2$, and
\begin{align}
a&=uA_{p,\downarrow}A_{q,\uparrow},\quad b=vA_{p,\uparrow}A_{q,\downarrow},\quad c=\epsilon\sqrt{1-\epsilon^2}\sqrt{A_{p,\uparrow}A_{p,\downarrow}A_{q,\uparrow}A_{q,\downarrow}},\\
d_1&=uA_{p,\downarrow}P_{q,\uparrow},\quad d_2=uP_{p,\downarrow}A_{q,\uparrow},\\
d_3&=uP_{p,\downarrow}P_{q,\uparrow}+vP_{p,\uparrow}P_{q,\downarrow},\\
d_4&=vA_{p,\uparrow}P_{q,\downarrow},\quad d_5=vP_{p,\uparrow}A_{q,\downarrow}.
\end{align}
In the ordered support basis $\{|\uparrow,\downarrow\rangle,|\downarrow,\uparrow\rangle,|\uparrow,\uparrow\downarrow\rangle,|\uparrow\downarrow,\downarrow\rangle,|\uparrow\downarrow,\uparrow\downarrow\rangle,|\downarrow,\uparrow\downarrow\rangle,|\uparrow\downarrow,\uparrow\rangle\}$, the reduced matrix is
\begin{equation}
\rho_{pq}=\begin{pmatrix}a&c&0&0&0&0&0\\c&b&0&0&0&0&0\\0&0&d_1&0&0&0&0\\0&0&0&d_2&0&0&0\\0&0&0&0&d_3&0&0\\0&0&0&0&0&d_4&0\\0&0&0&0&0&0&d_5\end{pmatrix}.
\label{eq:rhopq}
\end{equation}
Because $c^2=ab$, its nonzero eigenvalues are $\{a+b,d_1,d_2,d_3,d_4,d_5\}$. The marginal spectra are
\begin{align}
\Spec(\rho_p)&=\{uA_{p,\downarrow},vA_{p,\uparrow},uP_{p,\downarrow}+vP_{p,\uparrow},0\},\\
\Spec(\rho_q)&=\{uA_{q,\uparrow},vA_{q,\downarrow},uP_{q,\uparrow}+vP_{q,\downarrow},0\}.
\end{align}
Introducing $\HH(x_1,\ldots,x_m)=-\sum_i x_i\log_2x_i$, we obtain
\begin{align}
I(p:q)={}&\HH\big(uA_{p,\downarrow},vA_{p,\uparrow},uP_{p,\downarrow}+vP_{p,\uparrow}\big)\\
&+\HH\big(uA_{q,\uparrow},vA_{q,\downarrow},uP_{q,\uparrow}+vP_{q,\downarrow}\big)
-\HH(a+b,d_1,d_2,d_3,d_4,d_5).
\label{eq:Ipq}
\end{align}
The ordinary partial transpose produces one negative eigenvalue $-c$ and a corresponding positive eigenvalue $c$ in addition to the nonnegative diagonal part. Therefore,
\begin{equation}
{\LN(p:q)=\log_2\left[1+2\epsilon\sqrt{1-\epsilon^2}\prod_{r=p,q}\sqrt{A_{r,\uparrow}A_{r,\downarrow}}\right].}
\label{eq:LNpq}
\end{equation}

In the equal-spin limit $A_{r,\uparrow}=A_{r,\downarrow}=A_r$,
this expression becomes
$\log_2[1+2\epsilon\sqrt{1-\epsilon^2}A_pA_q]$
and reproduces the electric-only occupation calculation
of Ref.~\cite{Li2017}. For $\epsilon=1/\sqrt{2}$
and identical selected $p$ and $q$ modes,
Eq.~\eqref{eq:LNpq} reduces to
$\LN(p:q)=\log_2[1+A_\uparrow A_\downarrow]$.

For the remaining five bipartitions, the reduced
density matrices are obtained by tracing over
the complementary mode groups using
Eq.~\eqref{eq:traceformula}. We calculate the
corresponding mutual information and logarithmic
negativity numerically, without assuming equal
spin-dependent pair-production probabilities.
The analytical $p:q$ result and the electric-only
limits provide consistency checks. In particular,
the inter-group correlations and local mutual
information reproduce the corresponding results
of Ref.~\cite{Li2017}.

For the two local particle--antiparticle
partitions, the ordinary partial-transpose
spectrum gives
\begin{align}
\LN(p:-p)&=\log_2\!\left[
1+2\sqrt{
\epsilon^4 A_{p,\downarrow}P_{p,\downarrow}
+(1-\epsilon^2)^2 A_{p,\uparrow}P_{p,\uparrow}
}\right],
\label{eq:localLN1}\\
\LN(q:-q)&=\log_2\!\left[
1+2\sqrt{
\epsilon^4 A_{q,\uparrow}P_{q,\uparrow}
+(1-\epsilon^2)^2 A_{q,\downarrow}P_{q,\downarrow}
}\right].
\label{eq:localLN}
\end{align}

In the electric-only limit, these occupation-space
results differ from Eqs.~(30) and (32) of
Ref.~\cite{Li2017}: the negative eigenvalue depends
on the square root of the sum of squared coherences,
rather than their sum of squares. The reduced
density matrix and its ordinary partial-transpose
spectrum are derived in Appendix~\ref{app:audit}.
The comparison uses the same grouped-mode
occupation-space partition.

\subsection{Correlations at matched and different physical Landau levels}\label{sec:matchedstates}
Using the physical-level pairing established in Sect.~\ref{sec:matched}, let the particle groups occupy levels $N_p,N_q\geq1$. Define $P_r=P_{N_r}$ and $A_r=1-P_r$ for $r=p,q$, and set $u=\epsilon^2$. The resulting analytic pairwise correlations, obtained from Eq.~\eqref{eq:out} with $P_{r,\uparrow}=P_{r,\downarrow}=P_r$, are
\begin{align}
I(p:q)&=2A_pA_q h_2(u),&\LN(p:q)&=\log_2\left[1+2\sqrt{u(1-u)}A_pA_q\right],\label{eq:matched_pq}\\
I(p:-q)&=2A_pP_q h_2(u),&\LN(p:-q)&=\log_2\left[1+2\sqrt{u(1-u)}A_pP_q\right],\label{eq:matched_pminusq}\\
I(-p:q)&=2P_pA_q h_2(u),&\LN(-p:q)&=\log_2\left[1+2\sqrt{u(1-u)}P_pA_q\right],\label{eq:matched_minuspq}\\
I(-p:-q)&=2P_pP_q h_2(u),&\LN(-p:-q)&=\log_2\left[1+2\sqrt{u(1-u)}P_pP_q\right].
\label{eq:matchedinter}
\end{align}
To see explicitly how the first mutual-information identity arises, the particle marginals have spectra
\begin{align}
\Spec(\rho_p)&=\{uA_p,(1-u)A_p,P_p,0\},\\
\Spec(\rho_q)&=\{uA_q,(1-u)A_q,P_q,0\}.
\end{align}
Their entropies are $S(\rho_p)=h_2(P_p)+A_ph_2(u)$ and $S(\rho_q)=h_2(P_q)+A_qh_2(u)$. The nonzero eigenvalues of $\rho_{pq}$ are
\begin{equation}
\begin{aligned}
\Spec(\rho_{pq})=\{&A_pA_q,uA_pP_q,(1-u)A_pP_q,\\
&uP_pA_q,(1-u)P_pA_q,P_pP_q\},
\end{aligned}
\end{equation}
giving
\begin{equation}
S(\rho_{pq})=h_2(P_p)+h_2(P_q)+(A_pP_q+P_pA_q)h_2(u).
\end{equation}
Subtracting yields $I(p:q)=2A_pA_qh_2(u)$. The remaining inter-group formulas follow by retaining the other combinations of the mode-pair occupations, with the same coherent two-spin block. Direct diagonalization gives the respective partial-transpose trace norms $1+2\sqrt{u(1-u)}\,w$, where $w=A_pA_q,A_pP_q,P_pA_q,P_pP_q$ for the four inter-group partitions in the order stated.\par

The local correlations, in the same occupation and partial-transpose conventions, reduce to
\begin{align}
I(p:-p)&=2h_2(P_p),& \LN(p:-p)&=\log_2\left[1+2\sqrt{u^2+(1-u)^2}\sqrt{P_pA_p}\right],\label{eq:matchedlocal_p}\\
I(q:-q)&=2h_2(P_q),& \LN(q:-q)&=\log_2\left[1+2\sqrt{u^2+(1-u)^2}\sqrt{P_qA_q}\right].
\label{eq:matchedlocal}
\end{align}
The local mutual information is independent
of $u$, whereas the local logarithmic
negativity retains its dependence on $u$
for both equal and unequal physical levels.

The four inter-group mutual informations satisfy an exact redistribution identity,
\begin{equation}
I(p:q)+I(p:-q)+I(-p:q)+I(-p:-q)=2h_2(u)=I_{\mathrm{in}}(p:q).
\label{eq:matchedconservation}
\end{equation}

Following the pairwise-mode analysis of
Ref.~\cite{Li2017}, the four
inter-group logarithmic negativities satisfy
the nonlinear redistribution identity
\begin{equation}
\sum_{(X,Y)\in
\{(p,q),(p,-q),(-p,q),(-p,-q)\}}
\left[2^{\LN(X:Y)}-1\right]
=2\sqrt{u(1-u)}.
\label{eq:matchedLNidentity}
\end{equation}
This relation follows from the normalization
$A_r+P_r=1$ and retains the electric-only
structure of Eq.~(35) in Ref.~\cite{Li2017}. We next consider particle groups initially prepared at \emph{different} physical Landau levels, $N_p\neq N_q$, while the two spin representatives within each group are matched. In that case $P_p\neq P_q$ for $B>0$, even though each individual higher level remains spin degenerate. The two cross-group mutual informations in Eqs.~\eqref{eq:matched_pminusq} and \eqref{eq:matched_minuspq} need not coincide. For $0<u<1$ and $0<P_p,P_q<1$ their ratio is
\begin{equation}
\frac{I(p:-q)}{I(-p:q)}=\frac{A_pP_q}{P_pA_q}
=\frac{P_q/(1-P_q)}{P_p/(1-P_p)}.
\label{eq:crossratio}
\end{equation}
An identical ratio follows for the two negativity excesses $2^{\LN}-1$, rather than for the logarithmic negativities themselves:
\begin{equation}
\frac{2^{\LN(p:-q)}-1}{2^{\LN(-p:q)}-1}
=\frac{A_pP_q}{P_pA_q}.
\label{eq:crosslnratio}
\end{equation}
At fixed $E>0$, $B>0$ and otherwise equal mode labels, $N_q>N_p$ implies $P_q<P_p$ and therefore $I(p:-q)<I(-p:q)$; the same ordering holds for the two logarithmic negativities. For $N_p=N_q$ both cross partitions coincide. This difference between \emph{physical levels} is not an intrinsic spin splitting and does not survive when the corresponding probabilities become equal. The ratio therefore measures the asymmetric transfer of correlations between the selected levels. These relations follow from the pairwise-mode construction. Inserting the constant-field probability $P_N=\exp[-\pi(1+2N\mathcal B)/\mathcal E]$ shows explicitly that the ratio can be written as
\begin{equation}
\frac{I(p:-q)}{I(-p:q)}=\frac{\exp[\pi(1+2N_p\mathcal B)/\mathcal E]-1}{\exp[\pi(1+2N_q\mathcal B)/\mathcal E]-1},\qquad 0<u<1,\;\mathcal E>0,
\label{eq:cross_field_ratio}
\end{equation}
with the same expression for the ratio of negativity excesses. For $N_q>N_p$ and $\mathcal B>0$ this ratio is smaller than unity, whereas it equals unity at $\mathcal B=0$ within the equal-spin limiting model. This is a consequence of the unequal \emph{orbital gaps} of the prepared modes, not a Zeeman-type lifting of the higher-level spin degeneracy.

\begin{figure}
\centering\includegraphics[width=.80\linewidth]{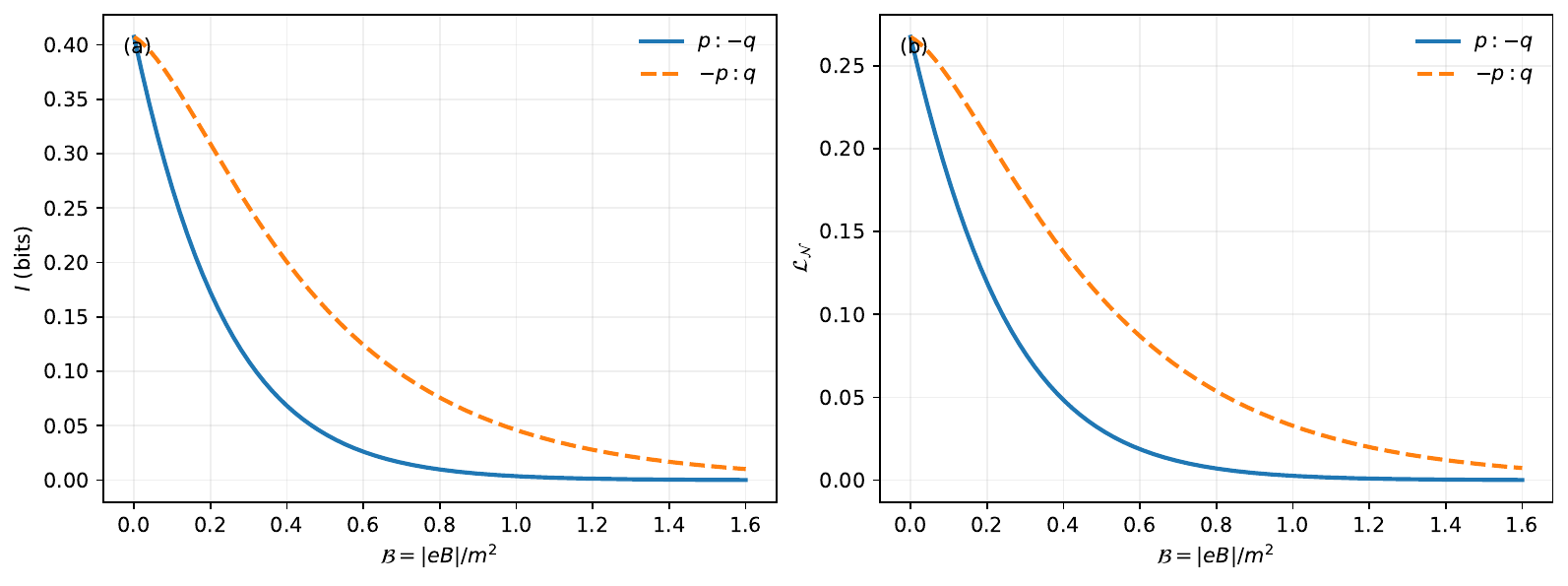}
\caption{\it \small Cross particle--antiparticle correlations for an initially entangled state prepared from different physical Landau levels $N_p=1$ and $N_q=2$ (the two spin representatives within each group remain properly matched). The mutual information $I$ (a) and logarithmic negativity $\LN$ (b) are plotted versus $\mathcal B=|eB|/m^2$ at fixed $\mathcal E=|eE|/m^2=2.5$ and $u=\epsilon^2=1/2$. Solid curves show $p:-q$, and dashed curves show $-p:q$. The correlations are equal at $B=0$ and become unequal when the magnetic transverse gaps differ, in agreement with Eqs.~\eqref{eq:crossratio} and \eqref{eq:crosslnratio}. The plots refer only to the selected grouped modes.}\label{fig:unequalcross}
\end{figure}

The redistribution identities retain the electric-only structure of Li, Dai and Shi~\cite{Li2017}, with $P_{N_r}$ accounting for the magnetic transverse gap. Figure~\ref{fig:unequalcross} illustrates this different-level asymmetry.

For $N_p=N_q=N$, the two cross
particle--antiparticle correlations
$(p:-q)$ and $(-p:q)$ depend on
$P_N(1-P_N)$ and attain their maxima
at $P_N=1/2$, provided this value is
accessible. At fixed
$\mathcal E=|eE|/m^2$, the corresponding
magnetic field is
\begin{equation}
\mathcal B_*=
\frac{(\mathcal E\ln 2)/\pi-1}{2N}.
\label{eq:Bstar}
\end{equation}

The cross particle--antiparticle correlations attain
an interior maximum at $\mathcal B_*>0$ for
$\mathcal E>\pi/\ln 2$; otherwise, their maximum
occurs at $\mathcal B=0$. In contrast, at fixed
$\mathcal E>0$ and $N>0$, the particle--particle
logarithmic negativity $\LN(p:q)$ increases
monotonically with $\mathcal B$ for $0<u<1$.
These results apply to selected Landau modes,
without summing over their degeneracy.

\begin{figure}[H]\centering\includegraphics[width=.75\linewidth]{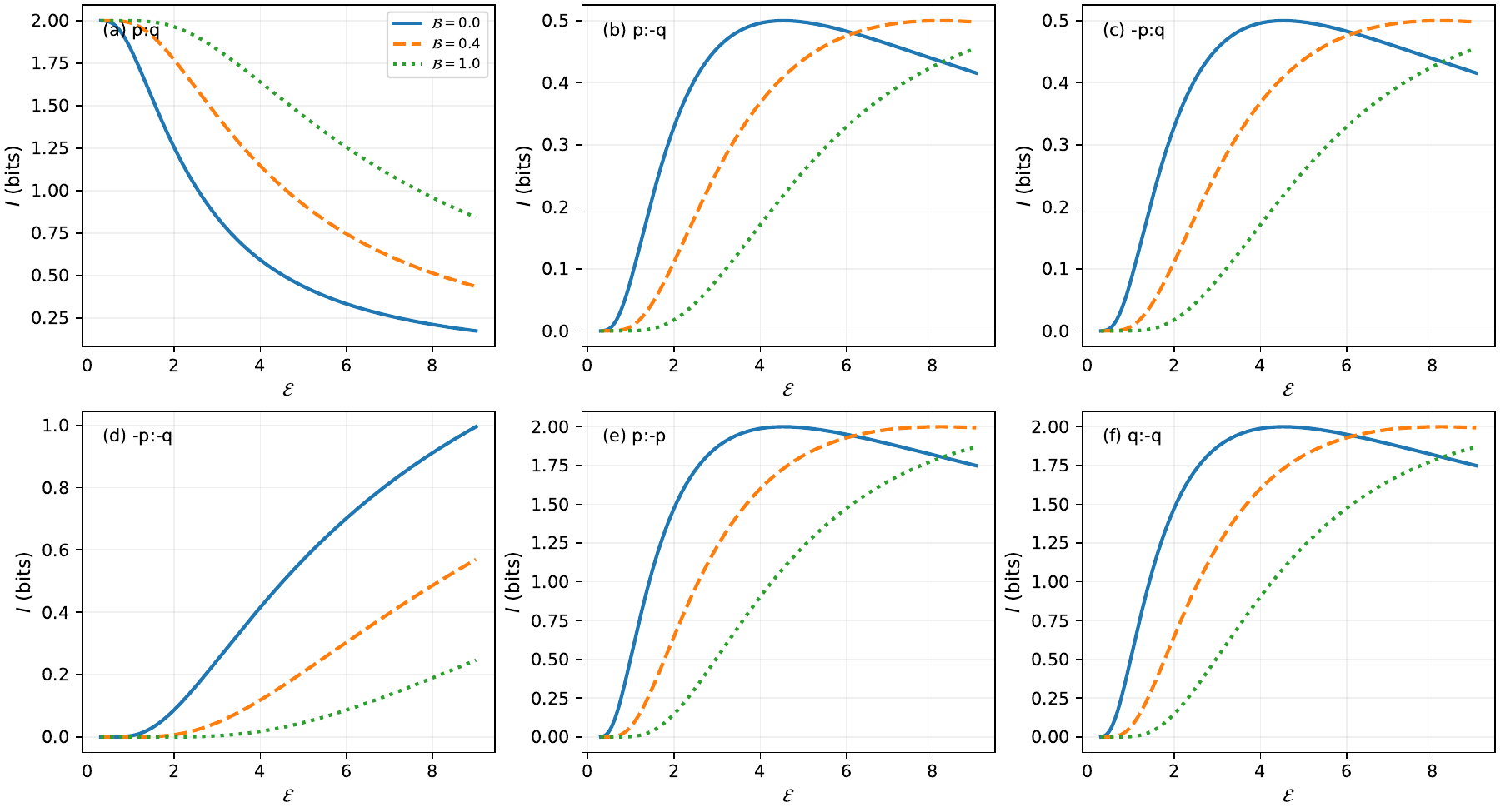}
\caption{\it \small Mutual information $I$ for the six paired grouped-mode partitions at the same physical Landau level $N_p=N_q=1$, corresponding to spin representatives $(n=0,\uparrow)$ and $(n=1,\downarrow)$ in each particle group. The horizontal axis is $\mathcal E=|eE|/m^2$; the three curves correspond to $\mathcal B=|eB|/m^2=0$, $0.4$ and $1$, with $u=\epsilon^2=1/2$. Panels (a)--(f) give $p:q$, $p:-q$, $-p:q$, $-p:-q$, $p:-p$ and $q:-q$. The plotted values follow Eqs.~\eqref{eq:matched_pq}--\eqref{eq:matchedinter} and \eqref{eq:matchedlocal_p}--\eqref{eq:matchedlocal} without a sum over Landau levels.}\label{fig:matchedMI}\end{figure}

\begin{figure}[H]\centering\includegraphics[width=.75\linewidth]{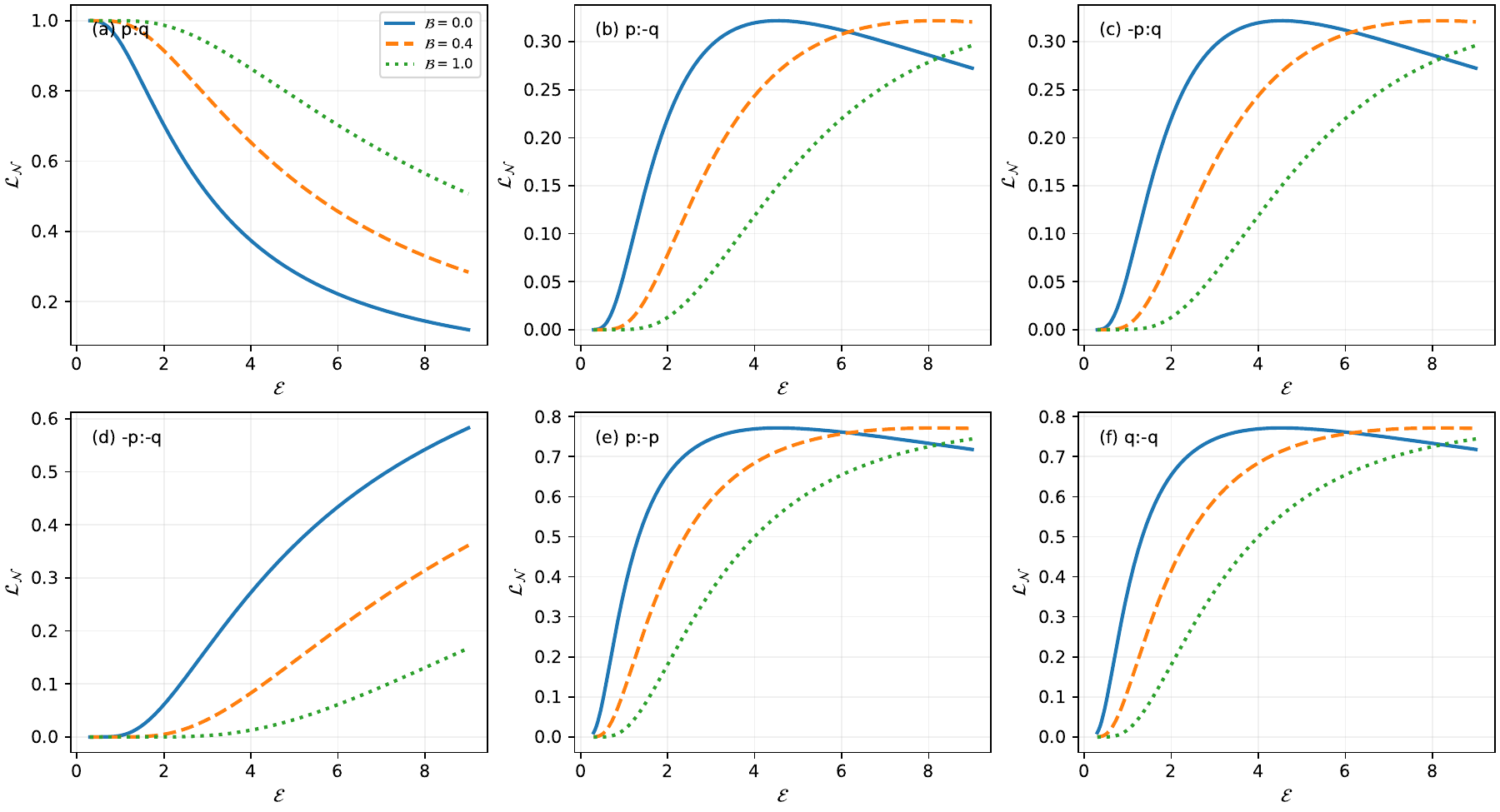}
\caption{\it \small Logarithmic negativity $\LN$ of the same six partitions for the same properly matched physical-level state and parameters as Fig.~\ref{fig:matchedMI}. The ordinary occupation-space trace norms are independently checked against the complete reduced density matrices in our numerical implementation. The local negativities employ the square-root coherence dependence in Eqs.~\eqref{eq:matchedlocal_p} and \eqref{eq:matchedlocal}.}\label{fig:matchedLN}\end{figure}

The matched-level results are shown in Figs.~\ref{fig:matchedMI} and \ref{fig:matchedLN}, while Fig.~\ref{fig:unequalcross} illustrates the unequal-level cross-correlation asymmetry.


\FloatBarrier
\section{Sauter-type electric field with a constant background magnetic field}
\label{sec:sauter}
\subsection{Pair-production probabilities and vacuum correlations}

We now consider a time-dependent electric field of the Sauter form \cite{Sauter1931,Adorno2017,Brout1991}
\begin{equation}
E(t)=E_0\operatorname{sech}^2(t/\tau),\qquad \mathbf B=B\hat{\mathbf z},
\label{eq:pulsefield}
\end{equation}
where $\tau$ characterizes the temporal scale of the Sauter-type electric field. For the choice $A_\mu=(0,-By,0,E_0\tau\tanh(t/\tau))$, the transverse oscillator equations remain unchanged, and the time-dependent Dirac equations can be solved in terms of hypergeometric functions \cite{Kaushal2025}. The in/out frequencies for a selected spin-resolved mode are
\begin{equation}
\omega_{\mathrm{in},s}=\sqrt{(k_z-eE_0\tau)^2+m^2+S_{n,s}},\qquad
\omega_{\mathrm{out},s}=\sqrt{(k_z+eE_0\tau)^2+m^2+S_{n,s}},
\label{eq:omegas}
\end{equation}
and $\varrho=|eE_0|\tau^2$. Denoting the Sauter-field probabilities by a superscript $\mathrm{S}$, the pair-production probability obtained in Eq.~(72) of Ref.~\cite{Kaushal2025} is
\begin{equation}
P^{\mathrm{S}}_{n,s}=\frac{\cosh(2\pi\varrho)-\cosh\!\left[\pi\tau(\omega_{\mathrm{in},s}-\omega_{\mathrm{out},s})\right]}{2\sinh(\pi\tau\omega_{\mathrm{out},s})\sinh(\pi\tau\omega_{\mathrm{in},s})},\qquad A^{\mathrm{S}}_{n,s}=1-P^{\mathrm{S}}_{n,s}.
\label{eq:pulseP}
\end{equation}
For real positive field amplitude and the chosen asymptotic mode convention, $0\le P^{\mathrm{S}}_{n,s}\le1$. At fixed $k_z$, these probabilities encode the dependence on $E_0$, $B$, $m$, $\tau$, spin and Landau index.

As the Sauter probabilities in Eq.~\eqref{eq:pulseP} depend on the transverse gap $S_{n,s}$, the same physical-level matching also applies to the Sauter-type electric field:
\begin{equation}
P_{N-1,\uparrow}^{\mathrm{S}}(k_z)=P_{N,\downarrow}^{\mathrm{S}}(k_z)\equiv P_N^{\mathrm{S}}(k_z),\qquad N\geq1.
\end{equation}
For two correctly matched spin branches, the analytic relations \eqref{eq:matched_pq}--\eqref{eq:matchedLNidentity} therefore remain true under the substitution $P_r\mapsto P_{N_r}^{\mathrm{S}}(k_{z,r})$. 

The pure two-mode-pair vacuum remains of the form in Eq.~\eqref{eq:vacstate}, with $P_s$ replaced by $P^{\mathrm{S}}_{n,s}$. The two measures therefore take the same algebraic form,
\begin{align}
I_{\mathrm{vac}}^{\mathrm{S}}&=2\sum_s h_2(P^{\mathrm{S}}_{n,s}),\\
\LN{}_{\mathrm{vac}}^{\mathrm{S}}&=2\sum_s\log_2\left(\sqrt{1-P^{\mathrm{S}}_{n,s}}+\sqrt{P^{\mathrm{S}}_{n,s}}\right).
\label{eq:pulsemeasures}
\end{align}

\begin{figure}
\centering\includegraphics[width=.98\linewidth]{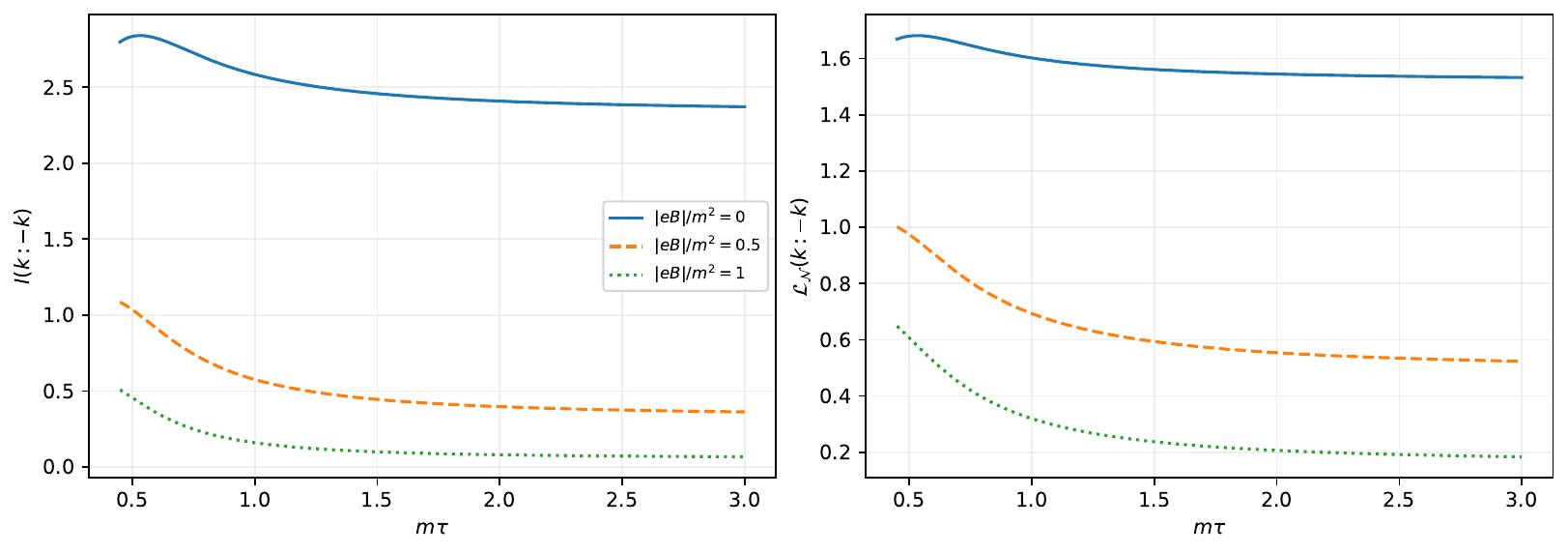}
\caption{\it \small Vacuum mutual information (left) and logarithmic negativity (right) versus dimensionless field time scale $m\tau$ for the Sauter electric field $E(t)=E_0\operatorname{sech}^2(t/\tau)$ with $|eE_0|/m^2=1.6$ and magnetic strengths $|eB|/m^2=0$, $0.5$, and $1$. The longitudinal momentum is $k_z/m=0$, the oscillator index is $n=1$, and both spin-resolved probabilities are calculated from Eq.~\eqref{eq:pulseP}.}
\label{fig:pulseline}\end{figure}

\subsection{Initially entangled states and correlation transfer}
The initially entangled state is transformed by the same fermionic mode-pair isometries with $P_{r,s}$ replaced by $P^{\mathrm{S}}_{r,s}$. In particular, the analytic $p:q$ expressions in Eqs.~\eqref{eq:Ipq} and \eqref{eq:LNpq} apply with these Sauter-field probabilities. We also evaluate all six grouped bipartitions from the full pure out-state by tracing out the complementary occupations; the graded permutation signs are retained. In the illustrative figures we take identical electric-field parameters for $p$ and $q$ and $k_{z,p}=k_{z,q}=0$, but the construction permits unequal longitudinal momenta. The phase-insensitive quantities considered here depend only on the magnitude of the Bogoliubov amplitudes for this specified state. The actual phases are important when more general coherent initial states or interferometric observables are considered.

\begin{figure}\centering\includegraphics[width=.99\linewidth]{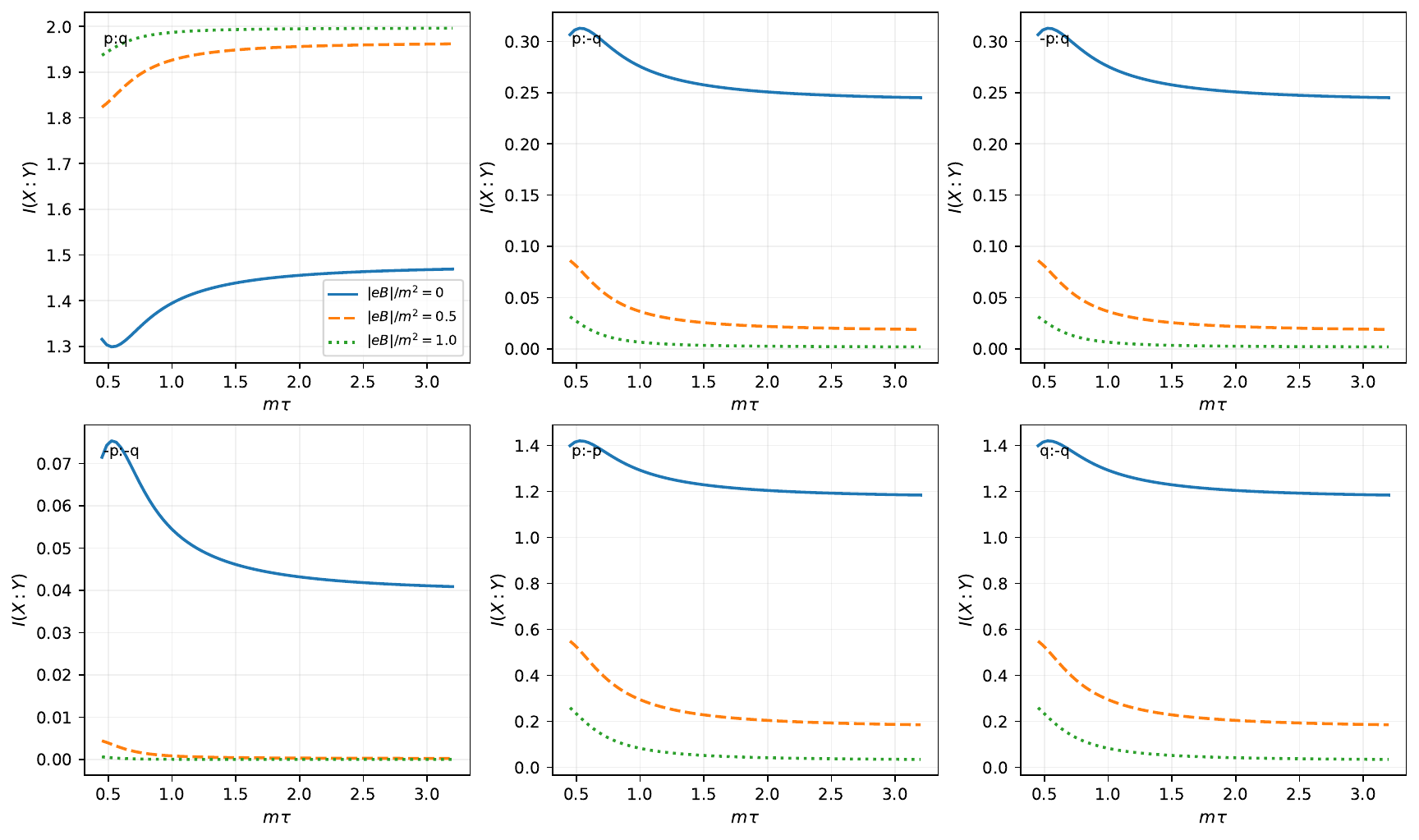}
\caption{\it \small Mutual information $I(X:Y)$ of the six initially entangled mode-group bipartitions as a function of dimensionless field time scale $m\tau$ for $E(t)=E_0\operatorname{sech}^2(t/\tau)$, $|eE_0|/m^2=1.6$, $|eB|/m^2=0,0.5,1$, $n_p=n_q=1$, $k_{z,p}=k_{z,q}=0$, and $\epsilon=1/\sqrt{2}$. In reading order the panels show $p:q$, $p:-q$, $-p:q$, $-p:-q$, $p:-p$, and $q:-q$. Every curve is computed from the full occupation-state density matrix with the Sauter-field probabilities in Eq.~\eqref{eq:pulseP}.}
\label{fig:pulseMI6}\end{figure}

\begin{figure}\centering\includegraphics[width=0.99\linewidth]{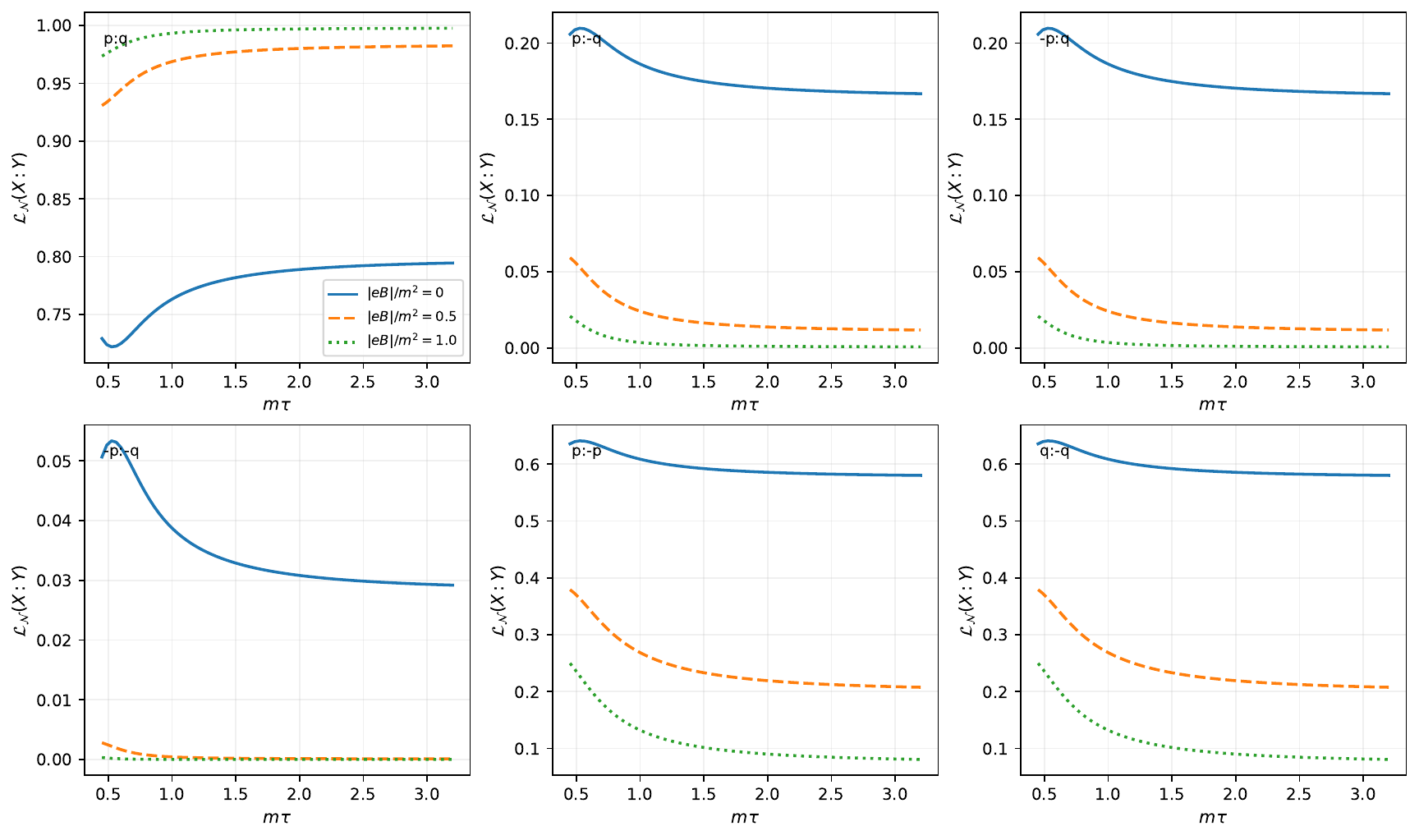}
\caption{\it \small Occupation-space logarithmic negativity $\mathcal L_{\mathcal N}(X:Y)$, defined by ordinary partial transposition, for the same six partitions, Sauter-field parameters, and initial state as Fig.~\ref{fig:pulseMI6}. 
The local particle--antiparticle panels
use the square-root coherence expressions
in Eqs.~\eqref{eq:localLN1} and \eqref{eq:localLN}, with the
corresponding Sauter-field probabilities.}
\label{fig:pulseLN6}\end{figure}

\begin{figure}\centering\includegraphics[width=1.0\linewidth]{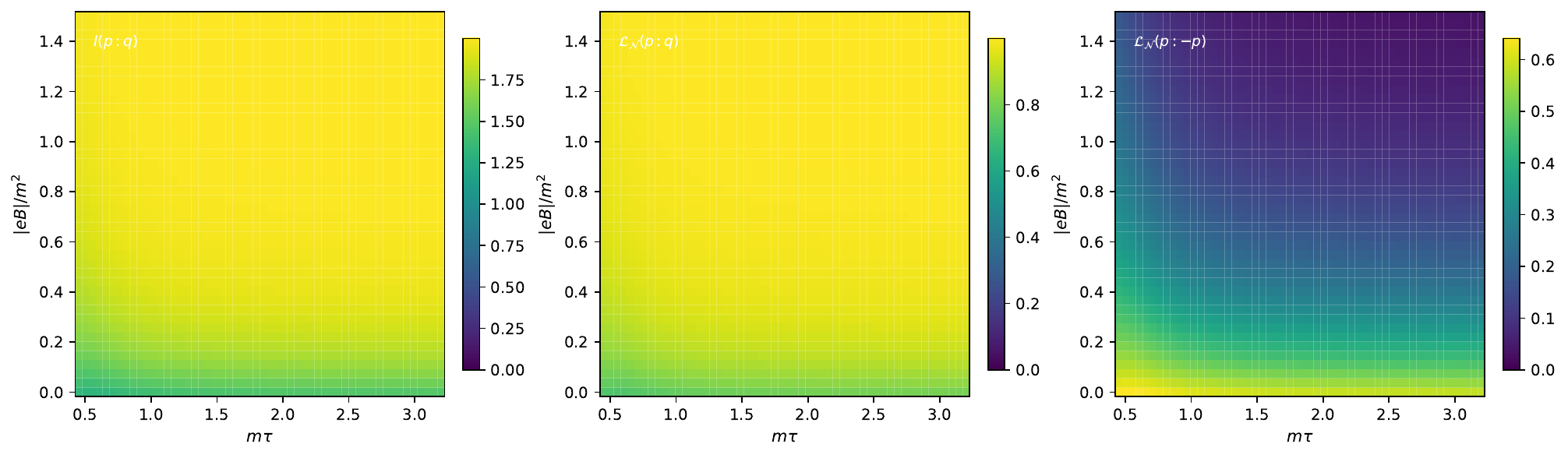}
\caption{\it \small Density maps of the mutual information $I(p:q)$, occupation-space logarithmic negativity $\mathcal L_{\mathcal N}(p:q)$, and local-pair logarithmic negativity $\mathcal L_{\mathcal N}(p:-p)$ for the initially entangled state in the plane $(m\tau,|eB|/m^2)$. We take $|eE_0|/m^2=1.6$, $n_p=n_q=1$, $k_{z,p}=k_{z,q}=0$, and $\epsilon=1/\sqrt2$. The color bars show the respective correlation measures; no Landau-level or guiding-centre sum is performed.}
\label{fig:pulsePairDensity}\end{figure}

For an initial state with $u=\epsilon^2$ and $v=1-u$, the particle--particle logarithmic negativity becomes
\begin{equation}
\mathcal L_{\mathcal N}^{\mathrm{S}}(p:q)=\log_2\!\left[1+2\sqrt{uv}\prod_{r=p,q}\sqrt{(1-P_{r,\uparrow}^{\mathrm{S}})(1-P_{r,\downarrow}^{\mathrm{S}})}\right].
\label{eq:pulseEntangledLN}
\end{equation}
The corresponding mutual information is given by Eq.~\eqref{eq:Ipq} with $A_{r,s}=1-P^{\mathrm{S}}_{r,s}$. The remaining inter-group and local partitions follow from the same reduced density matrices. Their logarithmic negativities are evaluated using the ordinary occupation-space partial transpose defined in Sec.~\ref{sec:setup}, with the fermionic creation-operator order kept fixed. The resulting mutual information and logarithmic negativity are plotted in Figs.~\ref{fig:pulseMI6} and \ref{fig:pulseLN6}, while Fig.~\ref{fig:pulsePairDensity} shows the corresponding density-map comparison.

At a fixed oscillator index, the two spin
sectors have different transverse gaps and
hence different pair-production probabilities.
For a constant electric field, increasing the
magnetic field suppresses pair production,
with a stronger suppression in the spin-up
sector. 
In the Sauter-type background, the dependence
on the magnetic field and characteristic time
scale is determined by Eq.~\eqref{eq:pulseP}.
Since the dependence of mutual information
and logarithmic negativity on the pair-production
probabilities varies with the chosen bipartition,
suppression of pair production does not
necessarily imply suppression of quantum
correlations.

The initially entangled particle--particle negativity in Eq.~\eqref{eq:LNpq} depends on the product of the no-pair probabilities $A_{r,\uparrow}A_{r,\downarrow}$. At fixed $E$ and selected $n$, increasing $B$ raises these factors, thereby retaining a larger fraction of the original $p:q$ entanglement. In contrast, correlations transferred to the antiparticle groups involve one or more pair-production probabilities $P_{r,s}$ and typically become smaller in the same regime. This is a statement about the chosen mode-resolved state and not a universal bound on the entanglement or information summed over an entire Landau band. The degeneracy of physical Landau levels scales with $|eB|$ and must be included before any statement about an integrated correlation density can be made.

For the Sauter-type electric field, the longitudinal frequencies entering Eq.~\eqref{eq:pulseP} retain the spin-dependent transverse gaps. Consequently the characteristic field time scale changes the two spin-resolved modes differently, and the full state can redistribute mutual information between particle--particle, cross particle--antiparticle, and local pairs in a manner that depends jointly on $m\tau$, $|eE_0|/m^2$, and $|eB|/m^2$. The Sauter-field plots show the corresponding dependence of the pairwise correlations on field strength and the characteristic field time scale. For large $m\tau$ and fixed $k_z$, the probabilities tend to the corresponding constant-field expressions, which we use as an independent numerical consistency check.

\FloatBarrier

\section{Conclusion}
\label{sec:conclusion}

In this work, we have investigated mutual information
and logarithmic negativity associated with fermionic
Schwinger pair production in parallel electric and
magnetic fields. Starting from the spin-dependent
Bogoliubov transformations, we constructed the
in-vacuum in terms of out-particle and antiparticle
occupation states and obtained the corresponding
reduced density matrices. The vacuum correlations
were derived analytically and depend on the
pair-production probabilities of the two independent
particle--antiparticle mode pairs.

We have distinguished the oscillator index from
the physical Landau-level index. Modes with the
same oscillator index generally have different
pair-production probabilities because they belong
to different physical Landau levels. In contrast,
the two spin representatives of a common higher
Landau level have identical transverse gaps and
pair-production probabilities, while the lowest
Landau level contains only one independent
spin-resolved mode. This distinction is essential
for interpreting the magnetic-field dependence
of the correlations.

For an initially entangled two-particle state,
we calculated mutual information and logarithmic
negativity for six particle--antiparticle
bipartitions. At matched physical Landau levels,
the inter-group correlations retain the
redistribution relations obtained in the
electric-only case, with the magnetic-field
dependence entering through the Landau-level
pair-production probabilities. When the particle
groups occupy different physical levels, the
two cross particle--antiparticle correlations
become unequal. We derived analytic relations
for this asymmetry and showed that it arises
from the difference in transverse gaps rather
than from spin splitting within a degenerate
physical Landau level.

We also derived the local particle--antiparticle
logarithmic negativity using the ordinary
partial transpose in a fixed occupation-space
basis. In the equal-probability limit, the
result differs from the corresponding expression
reported in Ref.~\cite{Li2017}. The difference
is explained through the partial-transpose
spectrum, and the limiting-case consistency
check is presented in Appendix B.

Furthermore, we extended the analysis to a
Sauter-type time-dependent electric field in
a constant magnetic background. The pairwise
correlations depend on the electric-field
strength, characteristic time scale, and
Landau-level structure. The constant-field
limit is recovered at large characteristic
time scales. Our results concern selected
fermionic modes and do not include summation
over Landau levels or their transverse
degeneracy.

The present framework can be extended to
more general time-dependent electric fields
and initially mixed fermionic states. Such
extensions would allow us to examine the
redistribution of quantum correlations
under more general field configurations
and initial conditions.

\section*{Acknowledgments}
S.~K. acknowledges financial support from Vellore Institute of Technology (VIT), Vellore, through the Faculty Seed Grant (RGEMS), Sanction Order No.~SG20260125.

\appendix
\numberwithin{equation}{section}
\section{Canonical pair-mode construction and fermionic signs}\label{app:operatorpair}

The occupation-state construction follows from the
diagonal fermionic Bogoliubov transformation.
In the normalized mode basis of
Ref.~\cite{Kaushal2025}, the overlap with the
antiparticle mode is diagonal in the mode index
$s$, and the corresponding
Bogoliubov coefficient determines the pair-production
probability $P_{r,s}=|\beta_{r,s}|^2$.
The label $\bar{s}$ denotes the paired antiparticle
mode and should not be identified with its physical
spin projection without specifying the
charge-conjugation convention.

For a single mode pair, suppress $r,s$, fix the creation-operator order $a_{\rm out}^{\dagger}$ before $b_{\rm out}^{\dagger}$, and take $\alpha=\sqrt{1-P}$ and $\beta=\sqrt P$. The canonical operators obey
\begin{equation}
a_{\rm in}=\alpha a_{\rm out}-\beta b_{\rm out}^{\dagger},\qquad
b_{\rm in}=\alpha b_{\rm out}+\beta a_{\rm out}^{\dagger},\qquad
\alpha^2+\beta^2=1.
\label{eq:canonicalcheck}
\end{equation}
The mixed anticommutators vanish, and $\{a_{\rm in},a_{\rm in}^{\dagger}\}=\{b_{\rm in},b_{\rm in}^{\dagger}\}=1$. Solving both annihilation conditions gives
\begin{equation}
|00\rangle_{\rm in}=\alpha|00\rangle_{\rm out}+\beta a_{\rm out}^{\dagger}b_{\rm out}^{\dagger}|00\rangle_{\rm out},\qquad
|10\rangle_{\rm in}=a_{\rm in}^{\dagger}|00\rangle_{\rm in}=|10\rangle_{\rm out}.
\label{eq:twomodeisometry}
\end{equation}
In checking the second equation one uses $b_{\rm out}a_{\rm out}^{\dagger}b_{\rm out}^{\dagger}|00\rangle=-a_{\rm out}^{\dagger}|00\rangle$. Thus the occupied fermionic mode cannot produce an additional identical particle, and the single-mode-pair map used in Eq.~\eqref{eq:iso} is a direct consequence of the canonical relations.

For two independent spin-resolved mode pairs we choose the grouped order
$a_{\uparrow}^{\dagger},a_{\downarrow}^{\dagger},b_{\bar\uparrow}^{\dagger},b_{\bar\downarrow}^{\dagger}$.
The product of two normalized pair vacua gives, in this order,
\begin{align}
|0\rangle_{\rm in}={}&\alpha_\uparrow\alpha_\downarrow|0\rangle_{\rm out}
+\beta_\uparrow\alpha_\downarrow a_\uparrow^{\dagger}b_{\bar\uparrow}^{\dagger}|0\rangle_{\rm out}
+\alpha_\uparrow\beta_\downarrow a_\downarrow^{\dagger}b_{\bar\downarrow}^{\dagger}|0\rangle_{\rm out}\nonumber\\
&-\beta_\uparrow\beta_\downarrow
 a_\uparrow^{\dagger}a_\downarrow^{\dagger}
 b_{\bar\uparrow}^{\dagger}b_{\bar\downarrow}^{\dagger}|0\rangle_{\rm out}.
\label{eq:fullpairvac}
\end{align}
The minus sign in the last term is required when $b_{\bar\uparrow}^{\dagger}$ is moved past $a_{\downarrow}^{\dagger}$. Its four squared amplitudes are exactly those used in Eq.~\eqref{eq:x}; their phases are not required for the vacuum Schmidt spectrum, but the operator-ordering sign is retained when constructing the initially entangled state and tracing nonadjacent modes. 
Equation~\eqref{eq:fullpairvac} describes the
selected particle--antiparticle mode pairs.
The antiparticle spin labels identify the
corresponding Bogoliubov partners and do not
necessarily represent their physical spin projections.

\section{Local particle--antiparticle reduced state and negativity}\label{app:audit}
We provide the density-matrix calculation underpinning Eq.~\eqref{eq:localLN1}. Let $u=\epsilon^2$, $v=1-\epsilon^2$ and write $A_s=1-P_s$. Tracing out both $q$-group modes leaves
\begin{equation}
\rho_{p,-p}=u|\psi_\uparrow\rangle\langle\psi_\uparrow|+v|\psi_\downarrow\rangle\langle\psi_\downarrow|,
\end{equation}
where phases can be absorbed consistently into the occupation basis and
\begin{align}
|\psi_\uparrow\rangle&=\sqrt{A_{p,\downarrow}}|\uparrow,0\rangle+\sqrt{P_{p,\downarrow}}|\uparrow\downarrow,\uparrow\rangle,\\
|\psi_\downarrow\rangle&=\sqrt{A_{p,\uparrow}}|\downarrow,0\rangle+\sqrt{P_{p,\uparrow}}|\uparrow\downarrow,\downarrow\rangle.
\end{align}
These two pure states are orthogonal and normalized. The density matrix has eigenvalues $u,v,0,\ldots,0$. Its two off-diagonal coherences have magnitudes
\begin{equation}
c_\downarrow=u\sqrt{A_{p,\downarrow}P_{p,\downarrow}},\qquad
c_\uparrow=v\sqrt{A_{p,\uparrow}P_{p,\uparrow}}.
\end{equation}
The ordinary partial transpose must be taken in the full $4\times4$ particle--antiparticle occupation basis. It contains, up to basis phases and permutation, the nontrivial block
\begin{equation}
M=\begin{pmatrix}0&c_\downarrow&c_\uparrow\\c_\downarrow&0&0\\c_\uparrow&0&0\end{pmatrix},
\qquad
\operatorname{spec}(M)=\left\{0,\pm\sqrt{c_\downarrow^2+c_\uparrow^2}\right\}.
\end{equation}
The remaining nonzero partial-transpose eigenvalues are $uA_{p,\downarrow}$, $uP_{p,\downarrow}$, $vA_{p,\uparrow}$, and $vP_{p,\uparrow}$. Hence
\begin{equation}
\|\rho_{p,-p}^{T_p}\|_1=1+2\sqrt{c_\downarrow^2+c_\uparrow^2},
\end{equation}

which gives Eq.~\eqref{eq:localLN1}. The $q:-q$ expression
in Eq.~\eqref{eq:localLN} follows by interchanging the
initially occupied spin assignments.
In the equal-spin limit, $A_\uparrow=A_\downarrow=A$ and $P_\uparrow=P_\downarrow=P$, we obtain
\begin{equation}
\LN(p:-p)=\log_2\left[1+2\sqrt{AP}\sqrt{\epsilon^4+(1-\epsilon^2)^2}\right].
\end{equation}

For $\epsilon=0$, the result reduces to
$\LN(p:-p)=\log_2[1+2\sqrt{AP}]$, giving
$\LN(p:-p)=1$ at $P=1/2$, as expected for
a maximally entangled two-level pure state.
The expression differs from Eq.~(30) of
Ref.~\cite{Li2017}, owing to the square-root
dependence obtained from the ordinary
partial-transpose spectrum.